# EPR Paradox and Magician's Props

Liangsuo Shu,[1,2] Shiping Jin,[1,2,3] Xiaokang Liu,[1] Suyi Huang,[1] Zhidong Zeng,[1] Alice Fox,[4] Kun Li[1,3] Jingyi Tan,[1,3]

[1]Department of Engineering Thermophysics, [2]Innovation Institute, [3]China-EU Institute for Clean and Renewable Energy, Huazhong University of Science & Technology, 1037 Luoyu Road, Wuhan, China
[4] Department of Chemistry, University of Oxford, Oxford, England

Local realism has been knocked down by the experiments with entangled pairs of particles based on Bell's theorem(J. S. Bell, Physics (Long Island City, N.Y.) 1, 195 (1964)). However, there has been continuing debate on whether locality or realism is the problem. In this work, we analyzed the Einstein-Podolsky-Rosen thought experiment of Bohm's version using information theory and thermodynamics. The inference of non-locality from EPR experiments will be against the principle of non-realism of quantum mechanics. Therefore, the experiments about quantum entanglement cannot provide any proof to accuse locality.

The EPR paradox[1] which used to be an imaginary weapon of Einstein to protect local realism by proving the incompleteness of quantum mechanics is now picked up to challenge the principle of locality conversely[2]. However, many other researchers including both experimental physicists and theoretical physicists questioned whether the experiments based on the Bell inequality provide ample evidence to convict locality[3]. In this work, we analyzed the EPR paradox of Bohm's version [4] using information theory and thermodynamics. We do not know whether it is good or bad news for Einstein: EPR paradox can neither protect the realism that he refused to abandon, nor bring any substantial danger to his cherished principle of locality.

There is a source of neutral pion $\pi^0$ that emits electron–positron pairs and the electron is sent to Alice while the positron is sent to Bob. They do this experiment under Prof. Charles's guidance. In quantum mechanics, it is meaningless to discuss any object without an observer. The two particles are in the entangled state. Each emitted pair occupies a quantum state: state 1 or state 2 (the spins at one axis, see Fig.1).

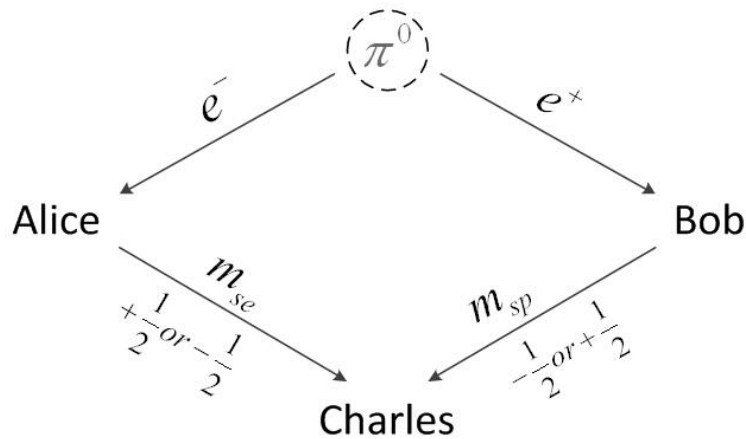

FIG.1 EPR paradox of Bohm's version

All of the information about the spin of the electron–positron pair (see Table. I) is

$$I_{\text{pair}} = -0.5 \times \ln 0.5 - 0.5 \times \ln 0.5 = \ln 2 \tag{1}$$

At time $t_1$ and $t_2$ ($t_2>t_1$) Alice and Bob measure the spin of electron and positron respectively and send their results to Prof. Charles through photons.

*Information processing of the measurement.*-Let us analyze the information processing by Alice first. Before her measurement, she knew that the electron spin quantum number $m_{se}$=+1/2 and $m_{se}$=-1/2 have the equal probability of 0.5.

$$I_{e1} = -0.5 \times \ln 0.5 \times 2 = \ln 2 \quad (2)$$

where $I_{e1}$ is the self-information of the electron before Alice's measurement.

**Table. I** Probability distributions of the spins of electron and positron pair

|           | electron | positron | probability |
|-----------|----------|----------|-------------|
| **State I**  | +1/2     | -1/2     | 0.5         |
| **State II** | -1/2     | +1/2     | 0.5         |

However, Alice can only get one result in one specific experiment: +1/2 or -1/2. After her measurement, the self-information of the electron will change into $I_{e2}$

$$I_{e2} = 1 \times \ln 1 = 0 \quad (3)$$

Therefore, Alice's measurement causes a decrease of the uncertainty of the electron (its self-information),

$$\Delta I_e = I_{e2} - I_{e1} = -\ln 2 \quad (4)$$

From equation 4, we can find that Alice's measurement is a process of entropy increment for the electron. The loss of information of the electron equals the information it has.

For Bob, before his measurement, he also faced the same difficulty as Alice to pick up one from two: the positron spin quantum numbers $m_{sp}$=+1/2 and $m_{sp}$=-1/2 have an equal probability of 0.5. The information from Alice is meaningless as it cannot help him to improve his forecast accuracy because he doesn't know the "information about the quantum entanglement of the particle-pair ($I_{qe}$)". Therefore, Bob's measurement also causes the same decrease of the self-information of the positron as Alice. The information of the positron decreases from $I_{p1} = \ln 2$ to $I_{p2} = 0$,

$$\Delta I_p = -\ln 2 \quad (5)$$

The loss of information of the electron should equal the information it has. However, for the electron–positron pair, the whole change in the information is

$$\Delta I_{pair} = \Delta I_e + \Delta I_p = -2\ln 2 \quad (6)$$

The measurements of Alice and Bob lead to a confusing result: the loss of information of the electron–positron pair is more than the information it has.

In fact, both Alice and Bob know nothing about quantum entanglement, the entangled electron–positron pair can be regarded as two independent particles (the probability distribution was shown in Table. II).

$$I_{pair}^* = -0.25 \times \ln 0.25 \times 4 = 2\ln 2 \quad (7)$$

Now, the information loss of the electron-positron pair equals the information it has. Their "ignorance" of $I_{qe}$ introduces additional pseudo information ($I_{a-p}=\ln2$) to the electron–positron pair unwittingly.

Charles gets information from Alice at $t_1+l/c$ and from Bob at $t_2+l/c$. After comparison, he finds that the spins of the electron and positron are different, which indicates that the information from Alice and Bob may be non-independent. What's more, the result would agree well with Tab.1 after the experiment is repeated enough times ($p(A=B)=0$).

**Table. II** Pseudo probability distributions of the spins of electron and positron

|     | electron | positron | probability |
| --- | --- | --- | --- |
| I   | +1/2 | +1/2 | 0.25 |
| II  | +1/2 | -1/2 | 0.25 |
| III | -1/2 | +1/2 | 0.25 |
| IV  | -1/2 | -1/2 | 0.25 |

After enough repeats of the experiment, Charles can speculate from the information he has received from Alice and Bob that the electrons and the positrons are entangled and maybe born in the decay of $\pi^0$. However in the next experiment, he still cannot predict Bob's information after he has received Alice's information, since every experiment is independent. At this time, it is entirely possible that the electron and the positron are not entangled and produced by other methods rather than the decay of $\pi^0$.

The researchers who reach the deduction of non-locality from EPR experiments assume subconsciously that the observer Charles already knew that the observed objects are entangled before the experiment. If Charles knows the prerequisite, he can get all of the information about the electron–positron pair before Bob's measurement. It seems that the measurement result from Bob (the spin of the positron) is decided by the measurement result from Alice. As $t_2$-$t_1$ can be an arbitrary positive number (under the premise of physics), does it mean there is some information which transfers faster than light (Einstein's 'spooky action at a distance') from Alice to Bob when $t_2$-$t_1$< $l_{ab}$/$c$? However, Charles cannot get the prerequisite he needs until he gets the results of both measurements from Alice and Bob. Then, he does not need to predict Bob's measurement result because he has already got it. Although the experiments based on Bell's inequality [4] have given a major blow to local realism, the spooky action at a distance is only a false appearance.

*Magician's props*-In a magic show, a white dove flies out from the tall hat of the magician. The hat is a well prepared prop, in which the white dove has already been hidden. The magician, who knows the existence of the white dove, would not feel surprised like his audiences (a popular magician usually pretends to make an exaggerated look of surprise). Any person, who has lost his (or her) precious childlike innocence, may admire the magician's perfect skill but not believe the "miracle". He knows that the dove was "stored" in the hat before and what the magician did was only to "read" it (free the dove) rather than to create it.

In an EPR experiment, the observer is the magician and the electron-positron pair is his prop. But he himself cannot determine whether the two particles are entangled or not until he receives their information. In other words, the magician does not know himself whether there is a white dove in his tall hat or not until he opens it. Before a measurement, any assumption about an entangled state is only a guess which needs to be confirmed. Otherwise, it will be against the principle of non-realism of quantum mechanics. Non-locality does not coexist with non-realism at least in the field of EPR experiments. We do not want to fall into the debate on whether quantum mechanics is non-local or not. Our work only indicates that the EPR paradox does not contradict the locality of relativity as well as the non-realism of quantum mechanics. Many outstanding physicists have made lots of meaningful progress on quantum entanglement[5] and are expected to make more in the future. However, they can not speculate any spooky action at a distance based on experiments about EPR paradox.

## ACKNOWLEDGMENTS

This work was partially supported by National Natural Science Foundation of China (No.51076057).